\documentclass[prl,aps,twocolumn,final]{revtex4-1} 
\usepackage{graphicx}
\usepackage{amsmath}
\usepackage{amssymb}
\usepackage{verbatim}

\newcommand{\beq}{\begin{equation}}
\newcommand{\eeq}{\end{equation}}
\newcommand{\ba}{\begin{array}{ccc}}
\newcommand{\ea}{\end{array}}

\newcommand{\Bamp}{{\mathcal{B}}}

\newcommand{\avbra}[1]{\left \langle #1 \right\rangle}
\newcommand{\bra}[1]{\left ( #1 \right)}
\newcommand{\sbra}[1]{\left [ #1 \right]}

\newcommand{\nearn}[2]{\avbra{#1,#2}}
\def\bea{\begin{eqnarray}}
\def\eea{\end{eqnarray}}
\def\be{\begin{equation}}
\def\ee{\end{equation}}

\begin{document}

\title{Fate of the Higgs mode near quantum criticality}

\author{Snir Gazit}
\author{Daniel Podolsky}
\author{Assa Auerbach}
\affiliation{Physics Department, Technion, 32000 Haifa, Israel}

\date{\today }

\begin{abstract}
We study a relativistic $O(N)$ model near the quantum critical point in 2+1 dimensions for $N=2$ and $N=3$. The scalar susceptibility is evaluated by  Monte Carlo simulation. We show that the spectrum contains a well defined peak associated with the Higgs mode arbitrarily close to the critical point. The peak fidelity and the amplitude  ratio  between the critical energy scales on both sides  of the transition are determined.
\end{abstract}

\maketitle
Spontaneously broken continuous symmetry in condensed matter produces collective modes.  In addition to Goldstone modes, an amplitude (Higgs) mode is sometimes  
expected at finite energy~\cite{VarmaHiggs,Huber:2007gv}. 
Higgs oscillations have been measured  in  e.g.,  the  superconductor NbSe$_2$, \cite{Sooryakumar:1980p809,LittlewoodVarma,VarmaHiggs}, the dimerized antiferromagnet TlCuCl$_3$ \cite{ruegg},  
and charge density wave compounds~\cite{ren,Pouget,yusupov}.

In the absence of gauge fields, the massive Higgs mode decays into massless Goldstone modes, broadening its spectral line. For relativistic $O(N)$ models in 3+1 dimensions, the Higgs mode becomes an increasingly sharper excitation the closer one gets to the quantum critical point (QCP)~ \cite{affleck_longitudinal_1992}.  This is a consequence of the fact that the QCP itself is a Gaussian fixed point.  In contrast, in 2+1 dimensions ($d=2$) the QCP is strongly coupled \cite{wf}, and there is no {\em a priori} reason to expect the Higgs mode to survive near criticality.

Recent  interest in the fate of the Higgs mode in 2+1 dimensions has led to new theoretical and experimental results. 
The visibility of the Higgs peak has been shown to be sensitive to the {\em symmetry} of the probe~\cite{lindner,PAA}:
The longitudinal susceptibility diverges at low frequencies as $\omega^{-1}$~\cite{ssrelax,zwerger,dupuisPRA}. This is due to the direct excitation of Goldstone modes which can completely conceal the Higgs peak.
In contrast, the  {\em scalar} susceptibility~\cite{PAA} rises as $\omega^3$ and its  Higgs peak is much more visible, even at stronger coupling. 

Indeed,  in recent experiments  
of cold bosons in an optical lattice~\cite{endres}, the Higgs mode  has been detected in the scalar response, in the vicinity of the superfluid to Mott insulator transition.
Further large $N$ analysis~\cite{dpss}, and numerical simulations of the Bose Hubbard model~\cite{pollet} (N=2),  have suggested that the Higgs peak is still visible as it softens toward the QCP. 
However the ultimate fate of this peak in the critical region demands simulations on much larger systems.

\begin{figure}
\includegraphics[scale=0.35]{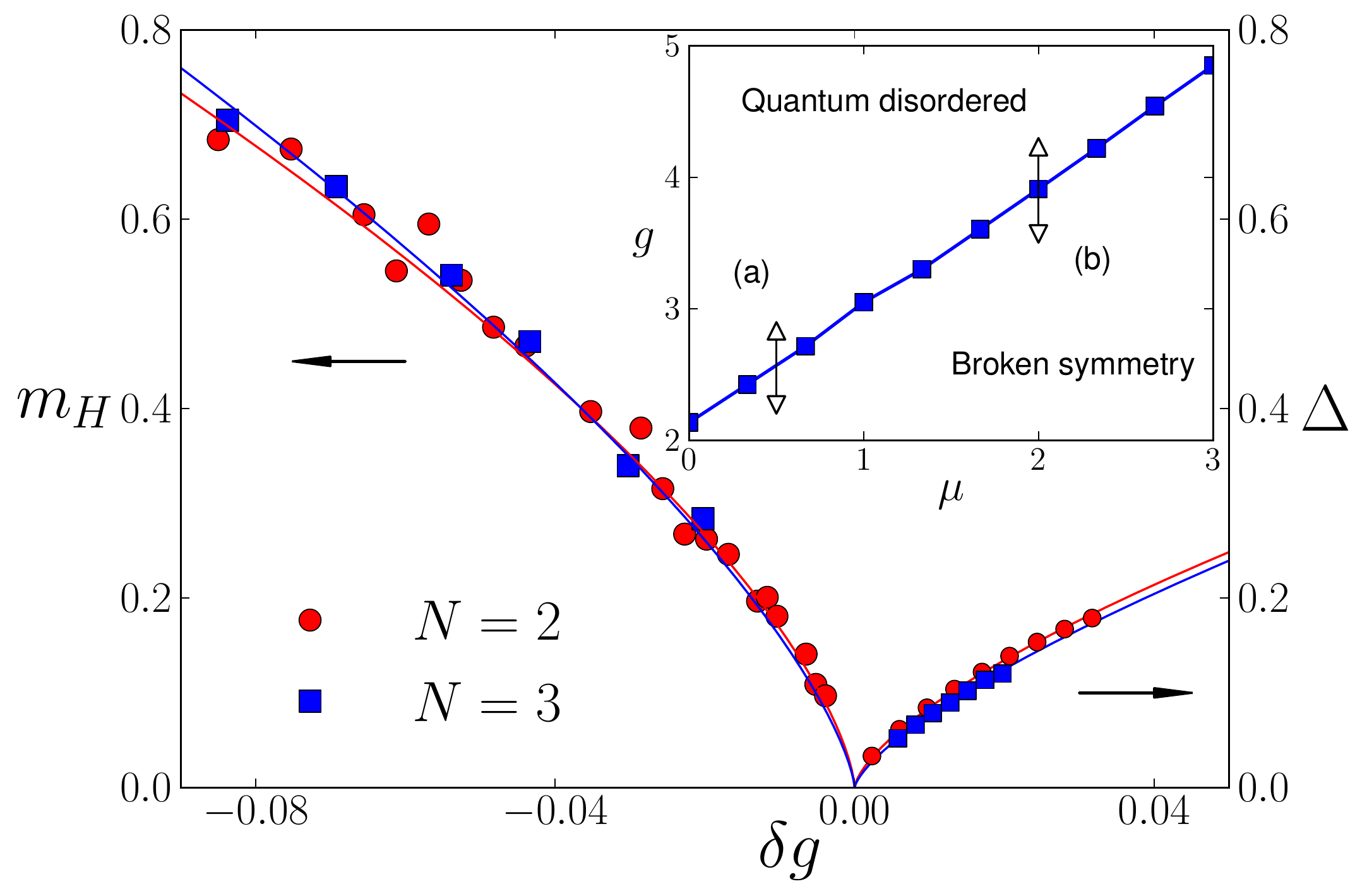}
\caption{Critical energy scales near the quantum phase transition in relativistic O(N) field theory for $N=2,3$.
$\delta g\equiv (g-g_c)/g_c$ is the
dimensionless tuning parameter. $m_H$ is the Higgs peak energy (mass) in the ordered phase $\delta g<0$, and $\Delta$ is the gap in the disordered phase $\delta g>0$.   
Solid lines describe  the critical behavior $m_H=\Bamp_- |\delta g|^{\nu_N}$ and  $\Delta=\Bamp_+ |\delta g|^{\nu_N}$.
Inset: Phase diagram in the microscopic $g, \mu$ parameter space (defined in Eq.~(\ref{eq:Z})  for $N=2$.   Two choices of $\mu$, denoted by (a) and (b), were studied in this paper. }
\label{fig:QCP}
\end{figure}

The question to be answered is whether the Higgs mode is well defined arbitrarily close to criticality, in the sense that one can identify a  
peak in the scalar susceptibility even as its energy scale decreases to zero.  

In this Letter we answer this question in the affirmative.   We  compute the frequency dependent scalar susceptibility for the relativistic O(2) and O(3)  models.
Lorentz invariance enables us to simulate large lattices and reach the close vicinity of the quantum critical point.

Our key  result is that in the broken symmetry phase, the scalar susceptibility collapses onto a universal lineshape. Its low frequency rise, and its peak, scale together with the vanishing Higgs mass $m_H$, which is plotted in Fig.~\ref{fig:QCP}.
The implication of our calculations, is that the {\em Higgs  mode remains a well defined collective mode all the way to the critical point}. 

We also determine  the amplitude ratios of the Higgs mass (in the ordered phase) to
the gap $\Delta$ (in the quantum disordered phase) for the O(2) and O(3) models, as $m_H/\Delta=2.1(3)$ and  $m_H/\Delta=2.2(3)$ respectively.
These ratios are universal quantities that can be directly compared with experiment.  Their value differs from the mean-field result $m_H/\Delta=\sqrt{2}$, which describes quantum critical points in $d=3$ for all values of $N$ \cite{QPTSubir}.


{\em Model --} We consider an $O(N)$ symmetric lattice model, with partition function $Z=\int {\mathcal{D}}\vec{\phi}\, e^{-S_E}$, where 
\begin{align}
S_E &= \frac 1 g \sbra{-\sum_{\nearn{i}{j}}\vec{\phi}_i\cdot\vec{\phi}_j-\mu\sum_{i}|\vec{\phi}_i|^2+\sum_{i}(|\vec{\phi}_i|^2)^2}.
\label{eq:Z}
\end{align}
Here, $\vec{\phi}_i$ is a $N$-component real field residing on  sites $i=(x,y,\tau)$ of a cubic lattice in discrete Euclidean space-time. The model undergoes a quantum phase transition at $g=g_c(\mu)$. See inset of Fig. 1. At weak coupling $\delta g=(g-g_c)/g_c<0$ there is long-range order, and the fluctuations include $N-1$ gapless Goldstone modes transverse to the broken symmetry direction $\langle \vec{\phi}\rangle$.  At strong coupling $\delta g>0$ there is a disordered phase with a gap $\Delta$ to all excitations.   Near the QCP, the long wave length properties of this model are captured by an $O(N)$ symmetric relativistic $\phi^4$ field theory in 2+1 dimensions \cite{QPTSubir}.  For $N=2$ the ordered  and disordered  phases describe the superfluid and Mott insulator of lattice bosons at commensurate filling, respectively~\cite{fisher_boson_1989}. For $N=3$, they describe the N\'eel ordered and the gapped singlet phase, respectively~\cite{Haldane_NonLinear_1983,CHN,QPTSubir}.

Our main focus is the zero-momentum scalar correlation function in imaginary time,
\bea
\chi_s\bra{\tau}&=&\frac{1}{L^2}\sum_{x,y}\left(\langle|\vec{\phi}_{(x,y,\tau)}|^2|\vec{\phi}_{\bf 0}|^2\rangle-\langle|\vec{\phi}_{\bf 0}|^2\rangle^2\right),
\label{eq:ScalarSusLat}\\
\tilde{\chi}_s(i\omega_m)&=&\frac{1}{L}\sum_{\tau} e^{-i \omega_m \tau } \chi_s\bra{\tau},
\label{eq:ScalarSusLat2}
\eea
and the real frequency dynamical susceptibility, given by
  \beq
\chi_s(\omega)=\tilde{\chi}_s(i\omega_m\to \omega+i0^+).
\label{eq:Wick}
\eeq

Scaling arguments indicate that, near $g_c$, the susceptibility at small frequencies is of the form \cite{dpss}
\be
\chi_s(\omega)\sim {C}+{\mathcal{A}}_\pm \Delta^{3-2/\nu} \Phi_{\pm}(\omega/\Delta),
\label{eq:ScalarSus}
\ee
Here, $\Delta\sim \Bamp_+ |\delta g|^\nu$ is the gap in the disordered phase, $\nu$ is the correlation length critical exponent, and $\Phi_-$ ($\Phi_+$) is a universal function of $\omega/\Delta$ in the ordered (disordered) side of the transition. The constant $C$ is real, and is a regular function of $g$ across the transition. The presence of Goldstone modes renders the ordered phase gapless.  In order to provide a well-defined energy scale that characterizes fluctuations on the ordered phase $\delta g<0$, we use the gap at the mirror point $-\delta g$ across the transition.   Our goal is to compute the universal scaling functions $\Phi_{\pm}$ and to extract a set of universal parameters that can be compared with experiment.


{\em Methods --}  The partition function in Eq.~(\ref{eq:Z}) is reformulated as a dual loop model, for the cases of $N=2$ and $N=3$~\cite{GPA2}. The sum is sampled by a Monte Carlo algorithm, 
using the  efficient ``worm algorithm'' \cite{prokofev_worm_2001}. An enhanced performance is achieved by cluster loop updates \cite{AletCluster2003}.  The simulations were performed on cubic  $L\times L \times L$ lattices of size up to $L=200$. This allowed us to approach within the neighborhood of $\delta g \ge 0.39\times10^{-2}$ of the QCP.

Our first task was to determine, for each set of parameters, the critical point $g_c$.
To this end we evaluated the helicity modulus $\rho_s$, as measured by the second moment of the winding number in periodic boundary conditions~\cite{prokofev_worm_2001,WangBiLayer2006}.
At $g_c$,  $\lim_{L\to \infty} L\rho_s(g_c)$ approaches  a universal value.  $g_c$ is then accurately determined from the crossing point of $L\rho_s$ for a sequence of  $L$ values~\cite{AletCluster2003}.  

The correlation length exponent $\nu$ is known from previous large scale simulations~\cite{hasenbusch_high-precision_1999,hasenbusch_eliminating_2001}:  
$\nu_2=0.6723(3)$ and $\nu_3=0.710(2)$ for $N=2$ and $N=3$ respectively.

For $N=2$, we study two sets of parameters $\mu_1=0.5$, for which  $g_c=2.568(2)$, and  $\mu_2=2$, for which $g_c=3.908(2)$, see inset of Fig. 1. 
Thus, the first set of parameters describes a ``softer'' spin model than the second.  For $N=3$, we use  $\mu=0.5$, for which $g_c=1.912(2)$.

{\em Results --}  We first extract the gap $\Delta$ in the  disordered phase at $\delta g>0$.
There, the imaginary part of the scalar susceptibility has a threshold at $2\Delta$ of the form \cite{dpss}, 
\bea
\Phi''_+(\omega/\Delta) \sim g_\gamma (\omega/\Delta) \Theta(\omega-2\Delta),
\label{eq:UniversalDis}
\eea
where $g_\gamma(x)=\pi/\left(\ln^2 \frac{x-2}{4\gamma}+\pi^2\right)$ has a weak logarithmic singularity and $\gamma$ is a universal constant which equals $1$ in the $N=\infty$ limit. 

The Laplace transform of Eq.~(\ref{eq:ScalarSus}) yields the asymptotics of large $\tau$ as
\be
\chi_s(\tau) \sim  {\mathcal{A}}_+ \tilde{g}_\gamma\bra{\tau \Delta} \Delta^{4-2/\nu}\frac{e^{-2 \tau\Delta}} {\tau\Delta} ,
\label{eq:CTauDis}
\ee
 where $\tilde{g}_\gamma\bra{\tau\Delta}$ is the Laplace transform of $g_\gamma (\omega/\Delta)$.
In order to obtain $\Delta$ we fit the simulation results for $\chi_s(\tau)$ to Eq.~(\ref{eq:CTauDis}).  The fit is dominated by the exponentially decaying factor, and is insensitive to the parameter $\gamma$.  
The critical behavior of the gap near the QCP agrees with the form
$\Delta= \mathcal{B}_+( \delta g)^\nu$,
as shown in Fig.~\ref{fig:QCP}.  From this procedure we extract the values of $\mathcal{A}_+$ and $\mathcal{B}_+$. As a check, we also obtained $\Delta$ from the large $\tau$ decay of the single-particle Green's function $G(k=0,\tau)$ \cite{Gap2DBose} and found agreement with these results.

\begin{figure}[b]
	\includegraphics[scale=0.4]{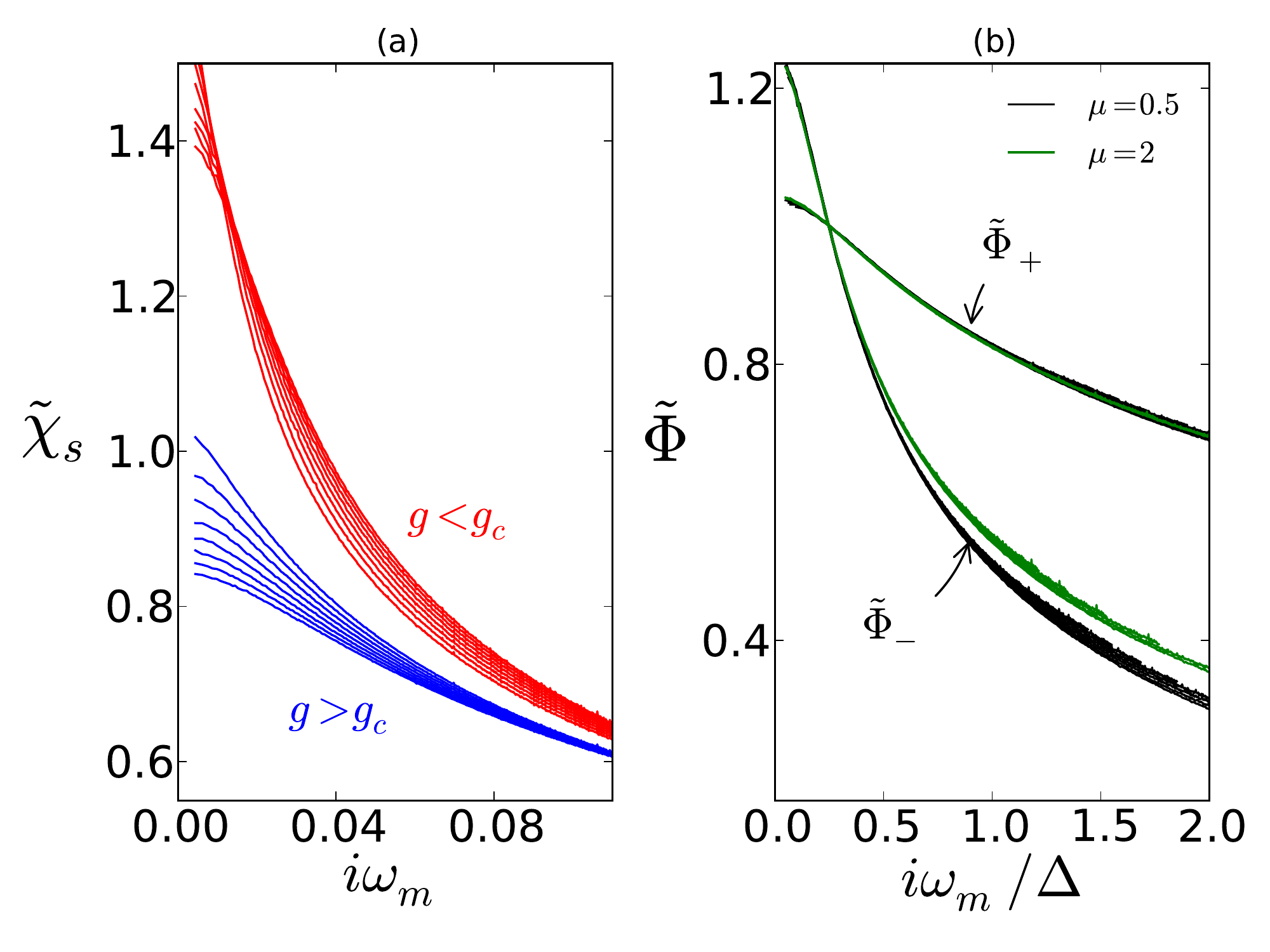}
\caption{(a) The scalar susceptibility as a function of Matsubara frequency for $N=2$ and $\mu=0.5$. Lower curves, different $\delta g>0$,  (disordered phase). Upper curves $\delta g<0$ (ordered phase) 
(b) Collapse of the same susceptibility onto  scaling  functions $\tilde{\Phi}_{\pm}$. Results include two values of $\mu$ (see inset of Fig.1)}
\label{fig:ImagFreqCol}
\end{figure}

The scaling form of Eq.~(\ref{eq:ScalarSus}) applies also to Matsubara frequency correlations $\tilde{\chi}_s(i \omega_m)$. 
Figure \ref{fig:ImagFreqCol}(a) shows the $N=2$ scalar susceptibility $\tilde{\chi}_s(i \omega_m)$ as a function of Matsubara frequency.
The data collapses into two curves corresponding to $\tilde{\Phi}_{\pm}(i\omega_m/\Delta)$ on both sides of the transition.  
Collapsing the curves is done by rescaling the Matsubara frequencies by $\Delta(\delta g)$, fitting the overall constant shift as a  polynomial of  $\delta g$, and then rescaling by $\mathcal{A}_+ \Delta^{3-2/\nu}$.  
The black curves in Fig.~\ref{fig:ImagFreqCol}(b) are the rescaled functions, which shows good convergence of the numerical data in the critical regime.
Similar results were obtained for $N=3$ and will be presented elsewhere\cite{GPA2}.

The universality of  $\tilde{\Phi}_{\pm}$ is tested in Fig.~\ref{fig:ImagFreqCol}(b). The  collapsed functions match closely, especially at  small Matsubara frequencies. We use  the values of $\mathcal{A}_+$ and $\mathcal{B}_+$ extracted in an earlier step, without free fitting parameters.  This provides a stringent test for the universality of the  scaling function.

\begin{figure}
\includegraphics[scale=0.4]{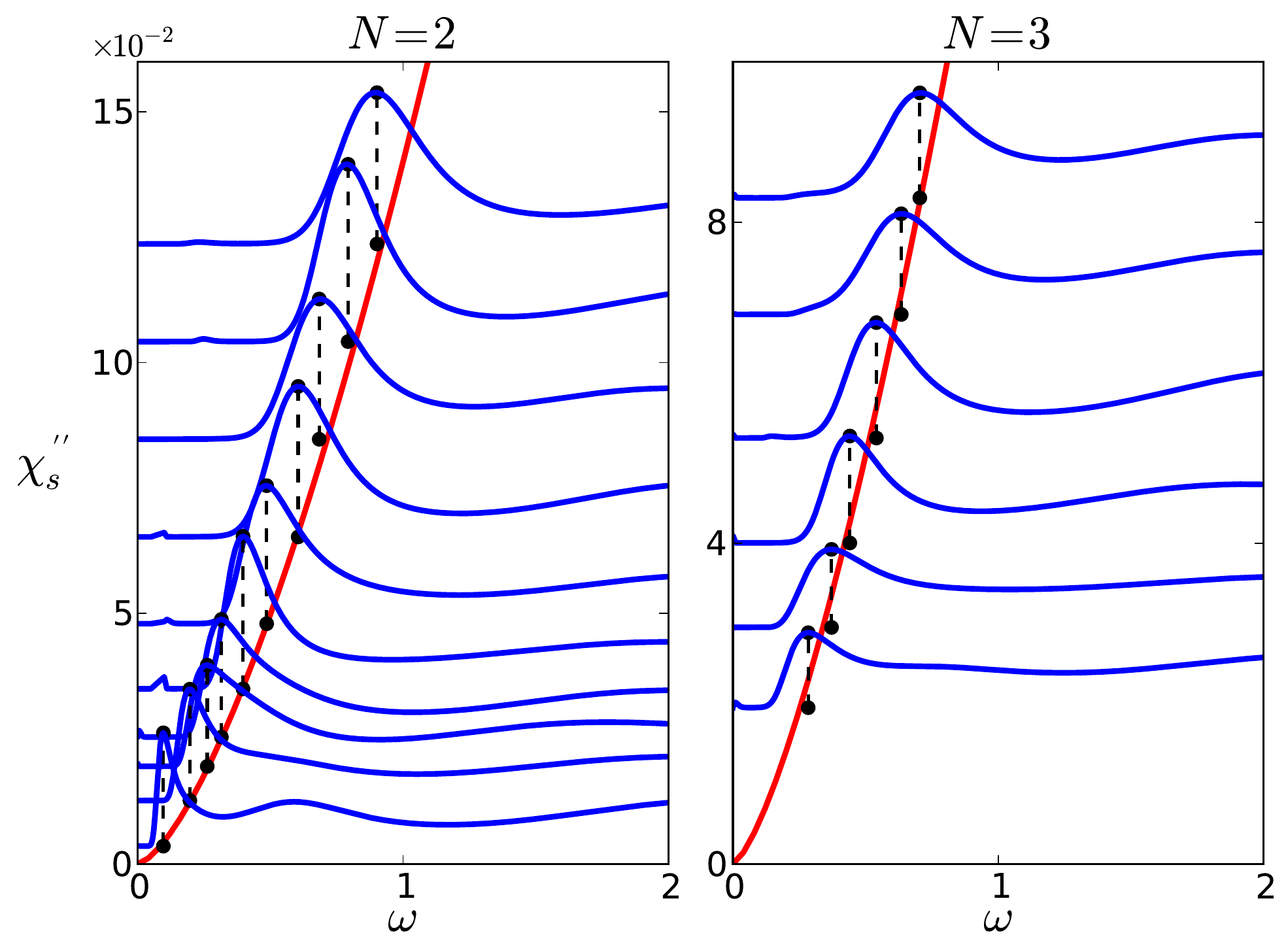}
\caption{
Scalar susceptibility $\chi_s''(\omega)$ (blue online), in arbitrary units, for different values of $\delta g<0$ at $\mu=0.5$. Curves shifted vertically by $\delta g$.  The Higgs energy $m_H$ as a function of $\delta g$ is extracted from the peak positions (dots) as the solid curve (red online).}

\label{fig:MaxEnt3D}
\end{figure}


{\em Analytic continuation --}
In order to perform the analytic continuation  of the numerical data, and determine the spectral function $\chi''_s(\omega)={\mathrm Im}\,\chi_s(\omega)$ we must invert the relation
\be
\tilde{\chi}_s(i\omega_m) = {1\over \pi} \int_{0}^{\infty}\! d\omega \chi''_s(\omega) \frac{2\omega}{\omega_m^2+\omega^2}.
\ee
Unfortunately the kernel of this integral equation is ill posed, which renders the inversion sensitive to inevitable numerical noise in $\tilde{\chi}_s(i \omega_m)$.
To tackle the problem we employ the MaxEnt Method~\cite{jarrell_bayesian_1996}. In this method the inversion kernel is regularized by introducing an ``entropy'' functional which is extremized along with the goodness of fit. To ensure the validity of the results we track the convergence of the MaxEnt spectrum as the statistical error decrease for long Monte Carlo simulations, see supplementary material\ref{sec:supp}.

Figure \ref{fig:MaxEnt3D} presents the spectral function $\chi''_s(\omega)$ in the ordered phase for $N=2$ and $N=3$. The spectral function displays a narrow low energy peak, which softens upon approach to the critical point, and broad high energy spectral weight, which does not.  This structure is in agreement with the findings of Ref.~\cite{pollet}.   
The position $m_H$ of the low energy peak as a function of $\delta g$ is shown 
in Fig.~\ref{fig:QCP}.  We find an excellent agreement with the expected scaling for the Higgs mode $m_H=\Bamp_- |\delta g|^\nu$ presented in the red curve.  From this fit we extract $\Bamp_-$.
From the ratio $\mathcal{B}_+/\mathcal{B}_-$ we extract the universal ratio  of the energy scales  $m_H/\Delta$ on both sides of the transition.

The $N=2$ results shown in Fig. \ref{fig:MaxEnt3D} range from $|\delta g|=0.0039$ to $|\delta g|=0.12$, corresponding to almost a decade and a half variation.  For the smallest value of  $\delta g$, the correlation length $\xi$ at the mirror point in the disordered phase is 22.2 lattice sites.  This value satisfies $1\ll\xi\ll L$, indicating that we are both in the continuum limit and in the thermodynamic limit.  For $N=3$, we observe scaling for a more narrow range of $|\delta g|$, between 0.02 to 0.084. In this case we find that approaching the critical point requires very large system sizes and long simulation times.  We will investigate smaller values of $\delta g$ for $N=3$ in a future study. For all cases presented here, we explicitly checked that our results do not change upon increasing $L$.

\begin{figure}[b]
\includegraphics[scale=0.4]{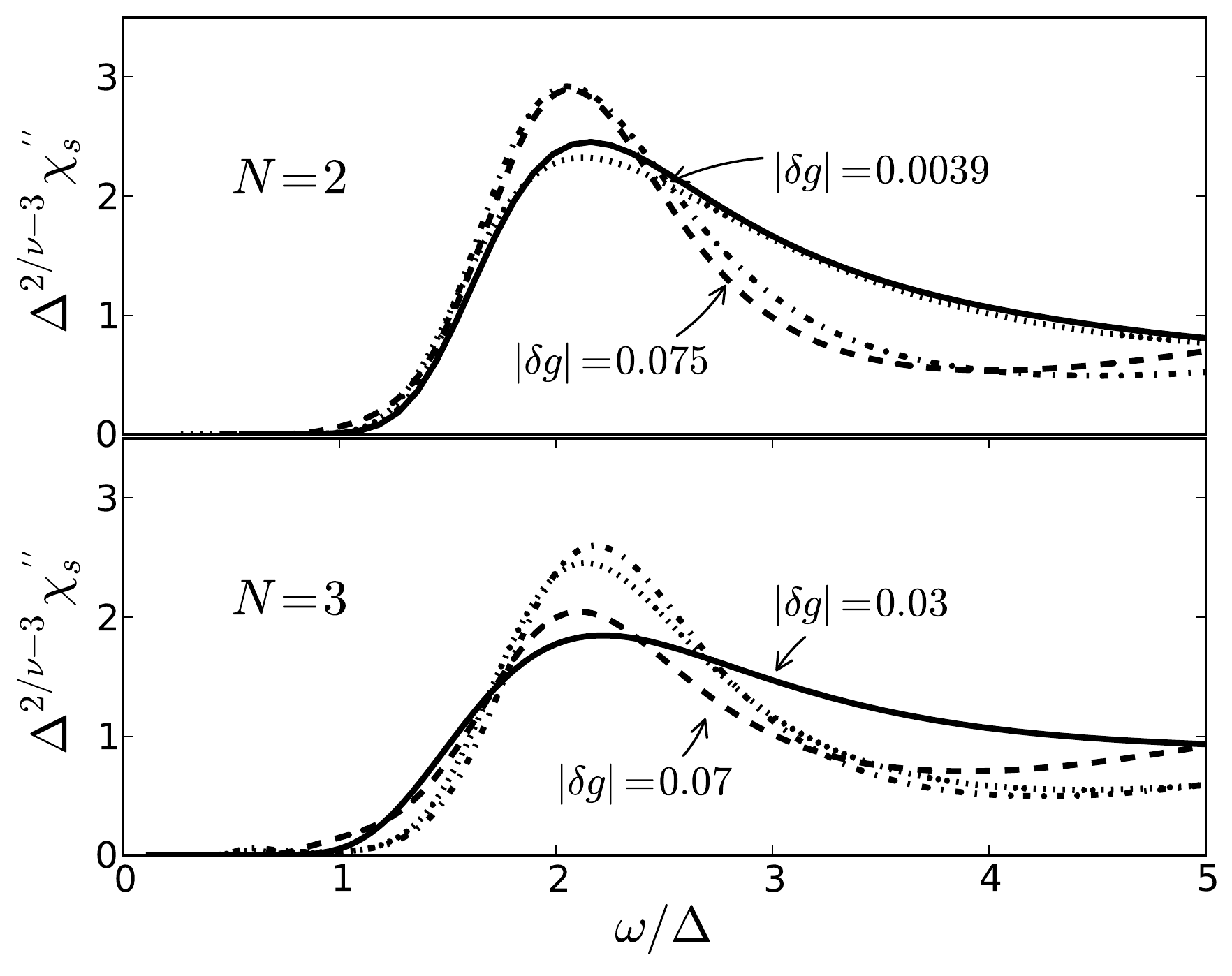}
\caption{
Rescaled spectral function vs.~$\omega/\Delta$ for $N=2,3$, at $\mu=0.5$.   At low values of $\omega/\Delta$, these curves collapse to the universal scaling function $\Phi''_-(\omega/\Delta)$, in accordance with Eq.~\ref{eq:ScalarSus}.    }
\label{fig:MaxEntSpectralExample}
\end{figure}

Not only does the peak position in Fig.~\ref{fig:MaxEnt3D} scale, but the full low energy functional form does, as shown in Fig.~\ref{fig:MaxEntSpectralExample} both for $N=2$ and $N=3$. There we rescale the frequency axis by $\Delta$ and the spectral function by $\Delta^{3-2/\nu}$ to match the predicted scaling form of Eq.~\eqref{eq:ScalarSus}. Note that the rescaling is done without any free fitting parameters since the real constant in Eq.~\eqref{eq:ScalarSus} drops out from the spectral function.  The observed functional scaling demonstrates that the Higgs peak is a universal feature in the spectral function that survives as a well-defined excitation arbitrarily close to the critical point. 

The peak position in units of $\Delta$ is shifted to higher energies for the $N=3$ case compared to $N=2$. This trend agrees with the prediction made in \cite{dpss} that $m_H/\Delta$ increases monotonically with $N$.  We also obtain the fidelity $F=m_H/\Gamma$, where $\Gamma$ is the full width at half-maximum. We measure $\Gamma$ with respect to the leading edge at low frequency, since at low frequencies there is less contamination from the high frequency non-universal spectral weight.
Since the entire functional form of the line shape is universal, $F$ is a universal constant that characterizes the shape of the peak. We find $F=2.4(10)$ for $N=2$ and  $F=2.2(10)$ for $N=3$. 

The rescaled spectral function in Fig.~\ref{fig:MaxEntSpectralExample} shows higher variability at high frequencies than at low frequencies. We attribute this to contamination from the non universal part of the spectrum and to systematic errors introduced from the MaxEnt analysis, which is less reliable in this regime.

In the ordered phase, the asymptotic low frequency rise of the susceptibility was predicted~\cite{ssrelax,PAA,dpss}  to be
\bea
\Phi''_- \sim \left(\omega/\Delta\right)^3,~~~  \omega\ll \Delta \ll 1.
\label{eq:UniversalOrd}
\eea
The $\omega^3$ rise  is due to the decay of a Higgs mode into a pair of Goldstone modes.  On the other hand,  Fig.~\ref{fig:MaxEntSpectralExample} does not display a clear $\omega^3$ low frequency tail.  An alternative method to look for this tail exists, without the need to 
analytically continue the numerical data to real time. Equation~(\ref{eq:UniversalOrd}) transforms into the large imaginary time asymptotics  $\chi_s\bra{\tau} \sim 1/\tau^{4}$. 

For $N=3$ we indeed find the asymptotic behavior $\chi_s(\tau)\sim 1/\tau^{4}$. Interestingly, for $N=2$ we do not find a conclusive asymptotic fall-off as  $1/\tau^4$,  Instead, the data fits better to an exponential decay, as in the disordered phase (see Eq.~\eqref{eq:CTauDis}). We find excellent agreement between the extracted decay rate and the value of $m_H$ obtained from the MaxEnt analysis, further validating our results for the Higgs mass. We note that the power law behaviour might be regained for larger values of $\tau$ below our statistical errors.
In both cases, we can safely conclude that the spectral weight of the Higgs peak dominates over the low frequency $\omega^3$ tail, enhancing its visibility.
The large $\tau$ analysis is discussed elsewhere \cite{GPA2}.

{\em  Discussion and Summary--}  Our results are directly applicable to all experimental probes that couple to a function of the order parameter {\em magnitude}.
For example, the lattice potential amplitude in the trapped bosons system~\cite{endres,pollet}, or pump-probe spectroscopy in Charge Density Wave systems \cite{ren,Pouget,yusupov}.
Such a probe can be expanded near criticality in terms of the order parameter fields  and their derivatives,
\bea
\Theta(x,\tau)=\alpha |\vec{\phi}|^2+\beta |\partial_\mu\vec{\phi}|^2+\gamma (|\vec{\phi}|^2)^2+\ldots
\eea
So long as $\alpha\ne 0$, the first term is more relevant than the rest.  
Hence, the scalar susceptibility defined in Eq.~(\ref{eq:ScalarSusLat}) dominates the experimental response at low frequencies and wave vectors.

In summary, we have calculated the scalar susceptibility for relativistic $O(2)$ and $O(3)$ models in 2+1 dimensions near criticality.
 We have demonstrated that the Higgs mode appears as a universal spectral feature surviving  all the way to the quantum critical point.  Since this is a strongly coupled fixed point, the existence of a well defined mode that is not protected by symmetry is an interesting, not obvious, result.
 We presented new universal quantities to be compared with experimental results.

During the submission of this paper we became aware of a similar analysis \cite{Kun} on the Bose-Hubbard model.

{\em Acknowledgements.--} We are very grateful to Nikolay Prokof’ev and Lode
Pollet for helpful comments and we also thank Dan
Arovas, Manuel Endres, Netanel Lindner, and Subir
Sachdev for helpful discussions. AA and DP acknowledge
support from the Israel Science Foundation, the European
Union under Grant No. 276923—MC-MOTIPROX, the
U.S.-Israel Binational Science Foundation, and thank the
Aspen Center for Physics, supported by the Grant
No. NSF-PHY-1066293, for its hospitality. S. G. received
support from a Clore Foundation Fellowship.

\appendix

\section{Supplementary Material for ``Fate of the Higgs mode near quantum criticality"}
\label{sec:supp}

Real frequency dynamics can be computed by analytic continuation of the imaginary time correlation function. Given the spectral function $A(\omega)$, the Matsubara Green's function $\mathcal{G}(i\omega_m)$ is obtained by
\beq
\mathcal{G}(i\omega_m)=\frac{1}{\pi}\int_0^\infty  \frac{2\omega }{\omega_m^2+\omega^2} A(\omega) d\omega
\eeq
(In the supplementary material section we use the notation ${G}(i\omega_m)$ and $A(\omega)$ to avoid confusion with the goodness of fit $\chi$.) Analytic continuation from Matsubara frequencies to real frequencies amounts to inversion of the problem above.

Unfortunately, the inverse of the kernel $K(i\omega_m,\omega)=\frac{1}{\pi} \frac{2\omega }{\omega_m^2+\omega^2}$ has exponentially growing singular values and is therefore ill-conditioned.  This renders the problem highly sensitive to the inevitable statistical noise in the Monte Carlo simulation. Naive minimization of the goodness of fit $\chi^2=(\mathcal{G}-K A)^T \Sigma (\mathcal{G}-K A)$, where $\Sigma$ is the covariance matrix, leads to an exponential amplification of the statistical noise obtained in $A(\omega)$.

To overcome this issue one must use a regularization procedure. One common approach is to introduce a cost function $f(A)$ that penalizes unphysical solutions, and to minimize the sum 
\beq
Q = \frac{1}{2} \chi^2+\lambda f(A)
\eeq
We will consider two such cost functions: (1) the maximum entropy choice $f(A)=-\sum_i A_i \log(A_i)$ \cite{MaxEnt} and (2) the Laplacian $f(A)=\sum_i \nabla^2 \log(A_i)$. Here $A_i$ refers to the values of the spectral function on a discretized frequency axis. The regularization parameter $\lambda$ is chosen so that the resulting spectral function is a good trade-off between the goodness of fit and the smoothing cost function. This is determined using the L-curve method \cite{LCurve} both for the MaxEnt and the Laplacian. In addition, for the MaxEnt we also use the ``classical maximum entropy'' \cite{MaxEnt} method based on a Bayesian statistics approach.  As a check of our methods, we verify at the end of our calculations that $\chi^2$ is close to the number of degrees of freedom, ${\cal N}$.  This ensures that our solution neither overfits nor underfits the statistical noise in the data.

Another approach we consider is the stochastic regularization \cite{Stoch1,Stoch2}. In this method the spectral function is obtained by averaging over a large sample of randomly-chosen solutions consistent with $\chi^2/{\cal N}\approx 1$. To produce such configurations we use the following procedure: First a random positive spectral function is generated. Then the goodness of fit is minimized using the steepest decent method while imposing positivity at each step. This procedure is repeated until $\chi^2/{\cal N} \approx 1$. Averaging over the random initial conditions leads to the final spectral function.

\begin{figure}[h!]
\includegraphics[scale=0.7]{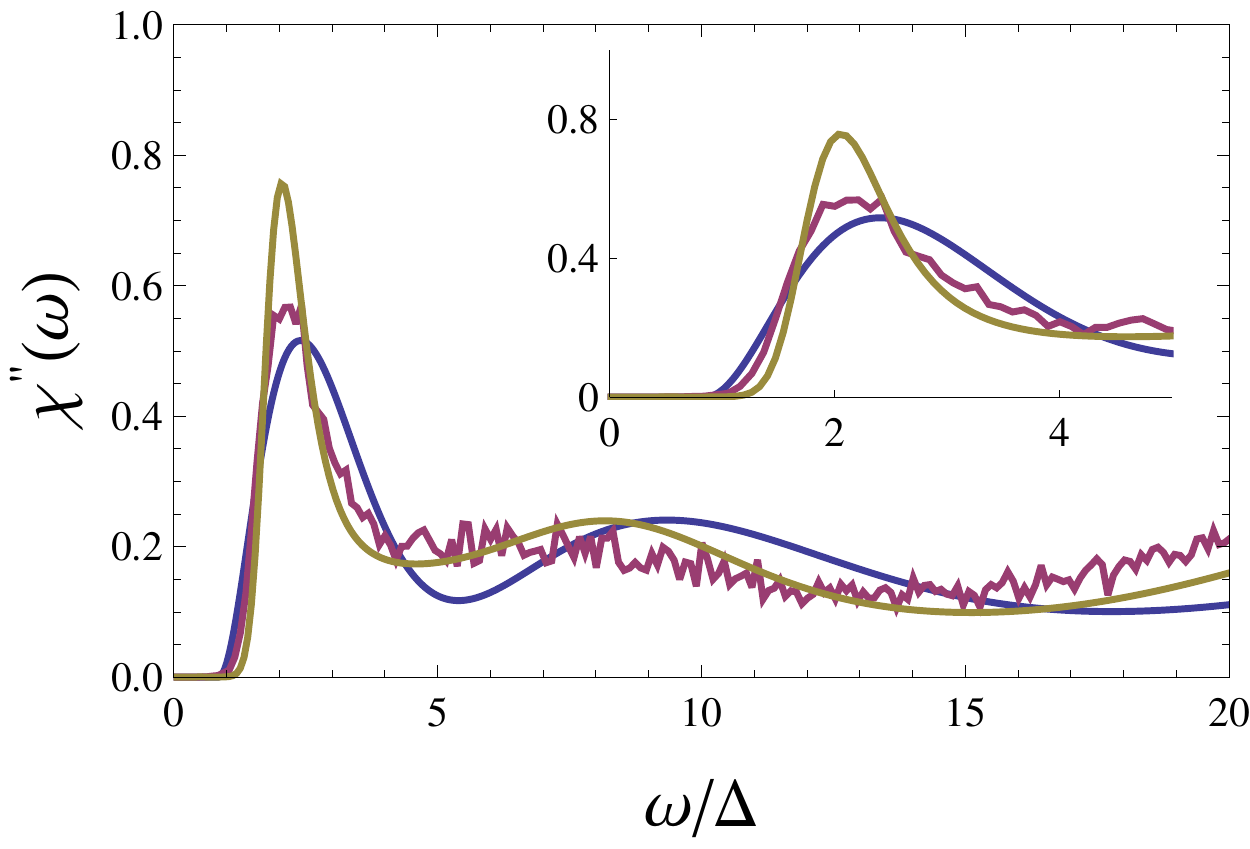}
\caption{$\chi_s^{\prime \prime}(\omega)$ obtained from different analytic continuation methods. MaxEnt (yellow), stochastic regularization (magenta), laplacian (blue). Inset focuses on the low frequency Higgs peak.}
\label{fig:sum_fig}
\end{figure}

As an example, we here present data for the $O(N=2)$ model with linear system size $L=120$ and detuning parameter $\delta g=1.17\%$. The results are summarized in Table \ref{tab:sum_tab} and the spectral functions are displayed in Fig.\ref{fig:sum_fig}.  Note that the position of the Higgs peak varies only slightly between different analytic continuation methods. 
\begin{table}[h!]
 \begin{center}
\begin{tabular}{ | c | c | c| c | }
\hline
  \bf{ Method }& \bf{$\bf{\chi^2}/{\cal N}$ }& \bf{$m_h/\Delta$} & \bf{Fidelity}\\
  \hline
  MaxEnt  & 0.984 & 2.03 & 2.36\\
  \hline
  Stochastic Regularization & $\approx 1$ & 2.1 & 1.75\\
  \hline
     Laplacian & 1.02 & 2.4  & 1.35\\
  \hline
\end{tabular}
\caption{Summary of results for different analytic continuation methods.}
\label{tab:sum_tab}
\end{center}

\end{table}


\bibliography{HIGGSMC}

\end{document}